\documentclass[journal]{IEEEtran}

\usepackage{xcolor,soul} %
\usepackage{siunitx}
\colorlet{shadecolor}{yellow}

\usepackage{graphicx}
\usepackage{epstopdf}
\graphicspath{{../pdf/}{../jpeg/}}
\DeclareGraphicsExtensions{.pdf,.jpeg,.png}
\newcolumntype{P}[1]{>{\centering\arraybackslash}p{#1}}
\usepackage{enumerate} 

\usepackage[cmex10]{amsmath}
\usepackage{amsfonts}
\usepackage{array}
\usepackage{url}
\usepackage{multicol, blindtext}
\usepackage{comment}
\usepackage[noadjust]{cite}
\usepackage{stmaryrd}
\usepackage[hidelinks]{hyperref}
\usepackage{float}
\usepackage{subcaption}

\usepackage{bm}

\usepackage[switch,columnwise]{lineno}

\usepackage{tikz,pgfplots}
\pgfplotsset{compat=newest}
\usetikzlibrary{plotmarks}
\usetikzlibrary{arrows.meta}
\usepgfplotslibrary{patchplots}
\usepackage{amsmath}
\usetikzlibrary{shapes,arrows,positioning,calc}
\usetikzlibrary{arrows.meta, calc, chains, positioning}
\usetikzlibrary{
	decorations.pathreplacing,
	shapes.symbols,
	shapes.geometric
}
\usetikzlibrary{backgrounds,
	intersections,
	positioning,
	quotes}

\DeclareMathOperator{\sign}{sign}

\begin{document}
\title{Probabilistic Shaping for High-Speed Unamplified IM/DD Systems with an O-Band EML}    	

\author{Md Sabbir-Bin Hossain,~\IEEEmembership{Graduate Student Member,~IEEE}, Georg B{\"o}cherer,~\IEEEmembership{Member,~IEEE},\\ Talha Rahman,~\IEEEmembership{Member,~IEEE}, Tom Wettlin, Neboj\v{s}a Stojanovi\'c,~\IEEEmembership{Member,~IEEE}, \\Stefano Calabr\`o, and Stephan Pachnicke,~\IEEEmembership{Senior Member,~IEEE}

 \thanks{Manuscript received 8 February 2023; revised 13 March 2023; accepted 24 March 2023. Date of publication XX XXXX 2023; date of current version XX XXXX 2023. The author(s) received no specific funding for this work. \textit{(corresponding author: Md Sabbir-Bin Hossain)}}
 \thanks{Md Sabbir-Bin Hossain is with Kiel University (CAU), Chair of Communications, Kaiserstr. 2, 24143 Kiel, Germany and with Huawei Munich Research Center, Riesstr. 25, 80992 Munich, Germany (e-mail: \hbox{sabbir.hossain@huawei.com}).}
 \thanks{ Georg B{\"o}cherer, Talha Rahman, Tom Wettlin, Neboj\v{s}a Stojanovi\'c, and Stefano Calabr\`o are with Huawei Munich Research Center, Riesstr. 25, 80992 Munich, Germany (e-mail:  \{georg.bocherer, talha.rahman, tom.jonas.wettlin, nebojsa.stojanovic, stefano.calabro\}@huawei.com).}%
 \thanks{Stephan Pachnicke  is with Kiel University (CAU), Chair of Communications, Kaiserstr. 2, 24143 Kiel, Germany (e-mail: stephan.pachnicke@tf.uni-kiel.de).}} 

\maketitle

\begin{abstract}
    Probabilistic constellation shaping has been used in long-haul optically amplified coherent systems for its
    capability to approach the Shannon limit and realize fine rate granularity. The availability of high-bandwidth
    optical-electronic components and the previously mentioned  advantages have invigorated researchers to explore
    probabilistic shaping (PS) in intensity-modulation and direct-detection (IM/DD) systems. This article
    presents an extensive comparison of uniform 8-ary pulse amplitude modulation (PAM) with 
    PS PAM-8 using cap and cup Maxwell-Boltzmann (MB) distributions as well as MB distributions of different Gaussian
    orders. We report that in the presence of linear equalization, PS-PAM-8 outperforms uniform PAM-8 in terms of
    bit error ratio, achievable information rate and operational net bit rate indicating that cap-shaped \hbox{PS-PAM-8} shows high
    tolerance against nonlinearities. In this paper, we have focused our investigations on O-band electro-absorption modulated laser unamplified IM/DD systems, which are operated close to the zero dispersion wavelength.
\end{abstract}

\IEEEpeerreviewmaketitle
\section{Introduction}\label{introduction}

\IEEEPARstart{C}{ommunication} channel models like the additive white Gaussian noise (AWGN) channel often have
non-uniform capacity-achieving input distributions. This has been the main motivation for probabilistic shaping (PS),
i.e., the development of practical transmission schemes that use non-uniform distributions at the input of the channel.
Many different PS schemes have been proposed in literature, see, e.g., the literature review in \cite[Sec.
II]{PAS_Georg}.

An important milestone for making PS practical was the invention of probabilistic amplitude shaping
(PAS)~\cite{PAS_Georg}, which concatenates a shaping outer code called a distribution
matcher~\cite{schulte2020algorithms} and a forward error correction (FEC) inner code. The PAS architecture has three
properties that distinguish it from other proposed PS schemes. First, it integrates PS with existing FEC, second, it
achieves the Shannon limit \cite[Sec. 10.3]{bocherer2018principles}, third, it supports configurability of the PS
overhead (OH) by changing the probability distribution. This additional degree of freedom is beneficial for transceiver
design. For instance, supposing that net bit rate, modulation format, and FEC OH are fixed and given, we can still
optimize a PAS design by trading PS OH against baud rate. Because of these three properties, PAS plays an important role
in coherent optical transmission systems~\cite{bocherer2019probabilistic} and state-of-the-art transceivers implement
PAS, see, for instance, \cite{li2018field,nokia2018pse3,sun2020800g}.
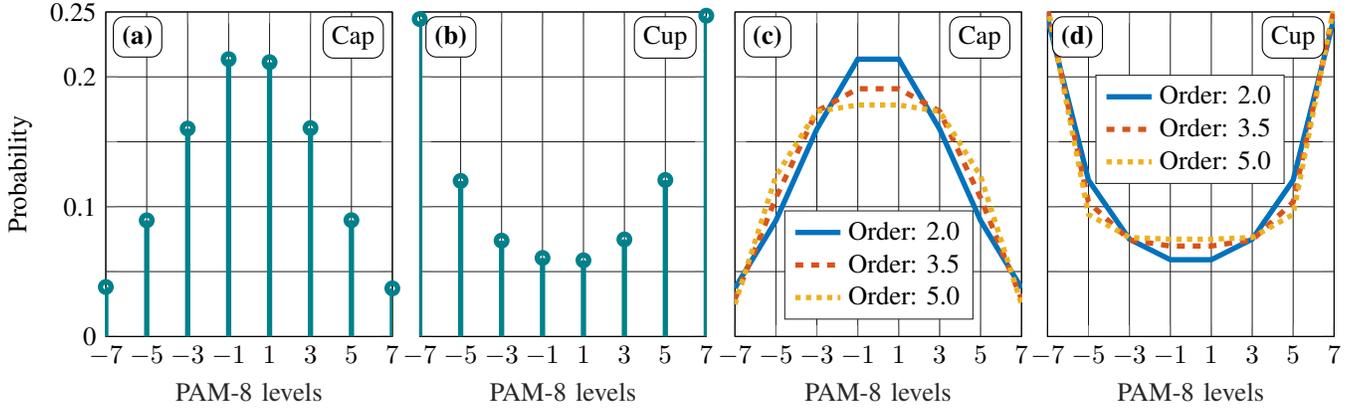
\begin{figure*}[t]
	\centering
%
%
\definecolor{mycolor1}{rgb}{0.00000,0.50588,0.54118}%
\definecolor{mycolor2}{rgb}{0.00000,0.44700,0.74100}%
\definecolor{mycolor3}{rgb}{0.85000,0.32500,0.09800}%
\definecolor{mycolor4}{rgb}{0.92900,0.69400,0.12500}%
\begin{tikzpicture}
	
	\begin{axis}[%
		width=1.5in,
		height=1.7in,
		at={(0.537in,0.532in)},
		scale only axis,
		xmin=-7,
		xmax=7,
		xtick={-7,-5,-3,-1,1,3,5,7},
		xlabel style={font=\color{white!15!black}},
		xlabel={PAM-8 levels},
		ymin=0,
		ymax=0.25,
		ytick={0,0.05,0.1,0.15,0.2,0.25},
		yticklabels={0,~,0.1,~,0.2,0.25},
		ylabel={Probability},
		axis background/.style={fill=white},
		xmajorgrids,
		ymajorgrids,
		grid style={line width=.1pt, draw=black!90},
		]
		\addplot[ycomb, color=mycolor1, line width=2.0pt, mark size=2.0pt, mark=o, mark options={solid, mycolor1}, forget plot] table[row sep=crcr] {%
			-7	0.0381480823863636\\
			-5	0.0894975142045455\\
			-3	0.160165127840909\\
			-1	0.213662997159091\\
			1	0.211310369318182\\
			3	0.160537997159091\\
			5	0.0896040482954545\\
			7	0.0370738636363636\\
		};
		
	\end{axis}
	
	\begin{axis}[%
		width=1.5in,
		height=1.7in,
		at={(2.180in,0.532in)},
		scale only axis,
		xmin=-7,
		xmax=7,
		xtick={-7,-5,-3,-1,1,3,5,7},
		xlabel style={font=\color{white!15!black}},
		xlabel={PAM-8 levels},
		ymin=0,
		ymax=0.25,
		ytick={0,0.05,0.1,0.15,0.2,0.25},
		yticklabels={\empty},
		axis background/.style={fill=white},
		xmajorgrids,
		ymajorgrids,
		grid style={line width=.1pt, draw=black!90},
		]
		\addplot[ycomb, color=mycolor1, line width=2.0pt, mark size=2.0pt, mark=o, mark options={solid, mycolor1}, forget plot] table[row sep=crcr] {%
			-7	0.244540127840909\\
			-5	0.119868607954545\\
			-3	0.0738547585227273\\
			-1	0.0605202414772727\\
			1	0.0587002840909091\\
			3	0.0748046875\\
			5	0.120578835227273\\
			7	0.247132457386364\\
		};
	\end{axis}
	
	\begin{axis}[%
		width=1.5in,
		height=1.7in,
		at={(3.830in,0.532in)},
		scale only axis,
		xmin=-7,
		xmax=7,
		xtick={-7,-5,-3,-1,1,3,5,7},
		xlabel style={font=\color{white!15!black}},
		xlabel={PAM-8 levels},
		ymin=0,
		ymax=0.25,
		ytick={0,0.05,0.1,0.15,0.2,0.25},
		yticklabels={\empty},
		axis background/.style={fill=white},
		xmajorgrids,
		ymajorgrids,
		grid style={line width=.1pt, draw=black!90},
		legend style={at={(0.172,0.052)}, anchor=south west, legend cell align=left, align=left, draw=white!15!black}
		]
		\addplot [color=mycolor2, line width=2.0pt]
		table[row sep=crcr]{%
			-7	0.0373233891759154\\
			-5	0.089296652671455\\
			-3	0.159736654198015\\
			-1	0.213643303954615\\
			1	0.213643303954615\\
			3	0.159736654198015\\
			5	0.089296652671455\\
			7	0.0373233891759154\\
		};
		\addlegendentry{Order: 2.0}
		
		\addplot [color=mycolor3, line width=2.0pt,dashed]
		table[row sep=crcr]{%
			-7	0.0288524209348239\\
			-5	0.106807782157545\\
			-3	0.173488467238529\\
			-1	0.190851329669103\\
			1	0.190851329669103\\
			3	0.173488467238529\\
			5	0.106807782157545\\
			7	0.0288524209348239\\
		};
		\addlegendentry{Order: 3.5}
		
		\addplot [color=mycolor4, line width=2.0pt,dotted]
		table[row sep=crcr]{%
			-7	0.0247592674415078\\
			-5	0.123553696619087\\
			-3	0.173343878417924\\
			-1	0.178343157521481\\
			1	0.178343157521481\\
			3	0.173343878417924\\
			5	0.123553696619087\\
			7	0.0247592674415078\\
		};
		\addlegendentry{Order: 5.0}
		
	\end{axis}
	
	\begin{axis}[%
		width=1.5in,
		height=1.7in,
		at={(5.466in,0.532in)},
		scale only axis,
		xmin=-7,
		xmax=7,
		xtick={-7,-5,-3,-1,1,3,5,7},
		xlabel style={font=\color{white!15!black}},
		xlabel={PAM-8 levels},
		ymin=0,
		ymax=0.25,
		ytick={0,0.05,0.1,0.15,0.2,0.25},
		yticklabels={\empty},
		axis background/.style={fill=white},
		xmajorgrids,
		ymajorgrids,
		grid style={line width=.1pt, draw=black!90},
		legend style={at={(0.167,0.47)}, anchor=south west, legend cell align=left, align=left, draw=white!15!black}
		]
		\addplot [color=mycolor2, line width=2.0pt]
		table[row sep=crcr]{%
			-7	0.24541297192515\\
			-5	0.120474476813984\\
			-3	0.0749710175035146\\
			-1	0.0591415337573511\\
			1	0.0591415337573511\\
			3	0.0749710175035146\\
			5	0.120474476813984\\
			7	0.24541297192515\\
		};
		\addlegendentry{Order: 2.0}
		
		\addplot [color=mycolor3, line width=2.0pt,dashed]
		table[row sep=crcr]{%
			-7	0.252417817609258\\
			-5	0.103501714701381\\
			-3	0.074379614089182\\
			-1	0.0697008536001789\\
			1	0.0697008536001789\\
			3	0.074379614089182\\
			5	0.103501714701381\\
			7	0.252417817609258\\
		};
		\addlegendentry{Order: 3.5}
		
		\addplot [color=mycolor4, line width=2.0pt,dotted]
		table[row sep=crcr]{%
			-7	0.254696421179155\\
			-5	0.0940809911818471\\
			-3	0.0762772322362723\\
			-1	0.0749453554027258\\
			1	0.0749453554027258\\
			3	0.0762772322362723\\
			5	0.0940809911818471\\
			7	0.254696421179155\\
		};
		\addlegendentry{Order: 5.0}
		
	\end{axis}
	
	\begin{axis}[%
		width=7.083in,
		height=1.7in,
		at={(0in,0.7in)},
		scale only axis,
		xmin=0,
		xmax=1,
		ymin=0,
		ymax=1,
		axis line style={draw=none},
		ticks=none,
		axis x line*=bottom,
		axis y line*=left
		]
		\node[fill=white, below right, align=left, draw=black,rounded corners]
		at (rel axis cs:0.236,0.888) {Cap};
		\node[fill=white, below right, align=left, draw=black,rounded corners]
		at (rel axis cs:0.699,0.888) {Cap};
		\node[fill=white, below right, align=left, draw=black,rounded corners]
		at (rel axis cs:0.466,0.888) {Cup};
		\node[fill=white, below right, align=left, draw=black,rounded corners]
		at (rel axis cs:0.93,0.888) {Cup};
		\node[fill=white, below right, align=left, draw=black,rounded corners]
		at (rel axis cs:0.081,0.888) {\textbf{(a)}};
		\node[fill=white, below right, align=left, draw=black,rounded corners]
		at (rel axis cs:0.312,0.888) {\textbf{(b)}};
		\node[fill=white, below right, align=left, draw=black,rounded corners]
		at (rel axis cs:0.775,0.888) {\textbf{(d)}};
		\node[fill=white, below right, align=left, draw=black,rounded corners]
		at (rel axis cs:0.547,0.888) {\textbf{(c)}};
	\end{axis}
\end{tikzpicture}%
	\caption{PS PAM-8 symbol distribution:
		(a) and (b) show the cap and cup-shaped variants of the MB distribution, respectively.
		Modified MB distributions with different Gaussian orders are shown for both cap and cup variants in (c) and (d),
		respectively.}
	\label{fig_PS_SymbolDistribution}
\end{figure*}

However, until today, for short-reach applications (up to 5~km), intensity-modulation with direct-detection (IM/DD) is
preferred over coherent transmission~\cite{IMDD_Xiang,IMDD_Jingchi}. Recent studies show progress towards $400$
Gb/s/lane~\cite{Berikaa:22_net400Gbps, Hossain_400G_ECOC21, Hossain_400G_OECC_PDP_21}, which leverage the early
availability of high bandwidth (BW) opto-electronic (OE) components. In this context, shaping gain and configurable PS
OH make PAS potentially attractive also for IM/DD systems~\cite{Hossain_PS_PAM_OFC21, Hossain_PS_PAM_ECOC21,
	Hossain_PS_PAM_ECOC22, Thomas_PS_PAM_JLT21, Che_AmplifierlessPAM_JLT21}.

The Maxwell-Boltzmann (MB) distribution, which is close to the optimum distribution for Gaussian
channels~\cite{Optimum_Kschischang}, minimizes the average power for a given source entropy. Since coherent systems
typically contain optical amplifiers and are subject to an average power constraint (APC), they benefit from the MB
distribution. However, typical IM/DD systems do not use optical amplification and, therefore, are considered unamplified
in this work (at high symbol rates IM/DD systems might require optical amplification). Unamplified IM/DD systems are
often considered as peak-power constrained (PPC). Under the idealized assumption that the modulator transfer function
has a sharp transition from the linear region to the saturation, it was shown that the cup (reverse) MB distribution is
beneficial and a shaping gain up to $0.47$ dB was demonstrated numerically for PAM-4~\cite{Thomas_PS_PAM_JLT21}. A
comprehensive study of probabilistic shaping (PS) for unamplified IM/DD was presented in
\cite{Che_AmplifierlessPAM_JLT21}, where both unipolar and bipolar MB and reverse MB distributions are investigated. The
key insight of \cite{Che_AmplifierlessPAM_JLT21} for unamplified IM/DD is as follows. If the system is PPC, PS does not
bring much gain. On the other hand, if the peak-to-average power ratio (PAPR) is already very large, e.g., due to
bandlimited components, then considerable PS gains can be observed.

In reality, the modulator transfer function shows a relatively slow transition from the linear to the nonlinear region.
This results in a new scenario in unamplified IM/DD systems, which lies somewhere in between PPC and APC systems. In
this paper, we discuss how to exploit the characteristics of an electro-absorption modulated laser (EML) by optimizing
the driving voltage both for 8-ary uniform and PS pulse-amplitude modulation (PAM).

This paper, which is supported by experimental demonstration with an unamplified IM/DD system, is a follow-up on our recently published
works~\cite{Hossain_PS_PAM_ECOC22, Hossain_PS_PAM_ECOC21, Hossain_PS_PAM_OFC21} and presents theoretical and physical
reasoning on the performance of multiple variants of the MB distribution.

The remainder of the paper is structured as follows. In Section \ref{sec_probabilistic_shaping}, PS is discussed for cap
(MB distribution) and cup-shaped (reverse MB distribution) variants with different Gaussian orders (GOs). Afterward,
driving voltage optimization and its performance are described in Section \ref{sec_DrivingVoltage_optimization}. Sections
\ref{sec_experimentalSetup} and \ref{sec_experimentalresult_discussion} describe the experimental setups used for
evaluating uniform and PS PAM-8 and present experimental results with discussions, respectively. Section
\ref{sec_conclusion} concludes the paper.
\section{Probabilistic Amplitude Shaping for IM/DD Systems} \label{sec_probabilistic_shaping}
\newcommand{\SE}{\textnormal{SE}}
\newcommand{\rfec}{R_{\textnormal{FEC}}}
\newcommand{\entop}{\textnormal{H}}
\newcommand{\cstllask}{\mathcal{X}_{\textnormal{ASK}}}
\newcommand{\cstllpam}{\mathcal{X}_{\textnormal{PAM}}}
\newcommand{\MB}{{\textnormal{MB}}}

PAS~\cite{PAS_Georg} concatenates at the transmitter side a distribution matcher (DM)~\cite{schulte2020algorithms} with
an off-the-shelf systematic FEC encoder. The DM imposes the desired distribution on the shaped bits, while the FEC
encoder generates additional parity bits. As no specific distribution can be imposed on the parity bits, they must be
treated as uniformly distributed.

\subsection{PAS for Coherent Transmission}
Coherent systems with optical amplification operate subject to an average power constraint. In this case the squared
amplitude of a symbol from the amplitude shift keying (ASK) constellation
\begin{align}
	\cstllask = \left\{\pm 1, \pm 3, \dotsc, \pm (2^m-1)\right\}\label{eq:ask constellation}
\end{align}
acts as the cost of the symbol. Consequently, PAS maps the shaped bits to amplitudes and imposes a distribution that minimizes the average power. The parity bits are mapped to signs, as the average power is invariant under the distribution of the signs. The distribution that minimizes the average power is an MB distribution given by
\begin{align}
	P_\MB^\nu(x) \propto e^{-\nu|x|^2},\quad \nu\geq 0, x\in\cstllask.\label{eq:MB distribution}		
\end{align}
The parameter $\nu$ controls the degree of shaping, i.e., for $\nu=0$, $X$ is uniformly distributed, while for $\nu$ large, the probability of the outer symbols approaches zero.

\subsection{PAS System Parameters~\cite{bocherer2019integration}}

For $2^m$-ASK constellations, there are $m-1$ amplitude bits and $1$ sign bit, so the FEC rate must be at least
$\frac{m-1}{m}$, since otherwise, the FEC encoder would produce more than $1$ parity bit per ASK symbol.
Correspondingly, the FEC OH must be smaller than $\frac{1}{m-1}$. For instance, for $16$-QAM, corresponding to $4$-ASK
in each real tributary, the FEC OH can be at most $100$\%, while for $64$-QAM, the FEC OH can be at most $50$\%. FEC OHs
of practical systems are smaller, so the OH constraint plays no further role in practice.

By \cite{bocherer2019integration}, the spectral efficiency of PAS with an ideal DM is
\begin{align}
	\SE = \left[\entop(X)-m(1-\rfec)\right]^+
\end{align}
where $\entop(X)$ is the entropy of the input in bits and where $[\cdot]^+=\max\{0, \cdot\}$ ensures the non-negativity
of $\SE$.

By \cite{bocherer2019integration}, \cite[Table~1]{bocherer2020probabilistic}, the PS redundancy is
$1-\frac{\entop(X)}{m}$ and the PS OH is
\begin{align}
	\textnormal{PS}_\textnormal{OH}= \frac{m}{\entop(X)}-1. \label{eq:PS_OH}
\end{align}
FEC and PS redundancy add up, so that the total OH is $\frac{1}{\rfec + \frac{\entop(X)}{m}-1}-1$.

\subsection{PAS for Square-Law Detection}

The work \cite{8346148} focuses on the square-law detection by a photodiode (PD) and assumes receiver AWGN added after
the PD. In this case, entropy subject to an average power constraint after the PD is maximized by an exponential
distribution before the PD. For the exponential distribution, PAS cannot be applied directly. In \cite{He:19}, a
pragmatic approach is proposed by assigning the same exponential distribution to the signal points with odd and even
index, respectively. Choosing between odd and even now plays the role of choosing the sign, and PAS can be applied.

\subsection{PAS for IM/DD}

In this work, we consider IM/DD without optical amplification and we aim to choose an input distribution suitable for the EML transfer curve, see Fig.~\ref{fig_EML_TransferCurve}. We choose a PAM constellation equal to the ASK constellation in \eqref{eq:ask constellation}, i.e.,
\begin{align}
	\cstllpam = \{\pm 1,\pm 3,\dotsc, \pm (2^m-1)\}\label{eq:pam constellation}
\end{align}
and we consider a more general family of MB distributions
\begin{align}
	P_{\MB}^{\nu, \alpha}(x)\propto e^{-\nu|x|^\alpha},\quad x\in\cstllpam\label{eq_Cap_Gaussian_order}
\end{align}
where $\alpha$ chooses the Gaussian order. The cost associated with symbol $x$ is now $\sign(\nu)|x|^\alpha$ and determined by the symbol amplitude so that when using the PAM constellation \eqref{eq:pam constellation}, the sign does not influence the average cost and PAS can be applied directly.

\subsubsection{Cap and Cup MB Distribution}

We set the Gaussian order to the common value $\alpha=2$. For cap-shaped distributions (Fig. \ref{fig_PS_SymbolDistribution}(a)) $\nu$ is positive, i.e., the exponent is
negative. Similarly, for cup-shaped MB distributions (Fig. \ref{fig_PS_SymbolDistribution}(b)), $\nu$ is negative,
i.e., the exponent is positive. As expected, both cap and cup MB distributions are symmetric around zero and, therefore, compatible
with PAS. The entropy of the cap and cup MB distributions can be set to the desired value by choosing $|\nu|$ accordingly, where a larger $|\nu|$ corresponds to a lower entropy. For each PS OH, there are only one possible cap-shaped and cup-shaped MB distribution, respectively.

\subsubsection{Different Gaussian Orders} \label{subsec_PS_Gaussian_order}

We now vary the Gaussian order (GO) $\alpha$~\cite{Hossain_PS_PAM_ECOC22}. A similar concept has been introduced in~\cite{Mohsen_SuperGaussian_ECOC2018} and resulted in super-Gaussian distributions for coherent systems. For each OH, we now have several candidate distributions depending on the order $\alpha$. The modification applies equally for the cap and cup-shaped variants.

The distributions with GO $2.0, 3.5~\textnormal{and}~5.0$ for cap and cup-shaped variants in the PAM-8 case are presented in Figs.~\ref{fig_PS_SymbolDistribution}(c) and (d), respectively. As the GO increases from $2.0$, corresponding to the classic MB
distribution, to GO $5.0$, the symbol occurrence of the center PAM levels becomes more uniform and the transition towards
the probability of the outer PAM levels becomes sharper. 
\section{Driving Voltage Optimization} \label{sec_DrivingVoltage_optimization}
In this section, we discuss the optimization of the peak-to-peak analog voltage ($V_{pp}$) as driving voltage of the modulator. As we
show in the following, this parameter controls the trade-off between extinction ratio (ER) and modulator nonlinearity
and affects the choice of the shaping distribution.

\begin{figure}[t]
	\centering
	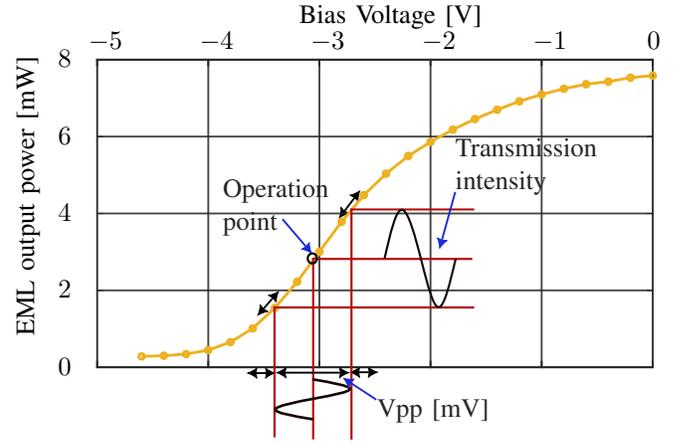
	\caption{EML transfer curve (measured) with exemplary modulation of a sine wave signal.}
	\label{fig_EML_TransferCurve}
\end{figure}

In our investigation we consider an EML operating near the zero-dispersion wavelength ($1310.9$ nm)
\cite{Hossain_400G_ECOC21, Hossain_400G_OECC_PDP_21}. The transfer curve of the EML, which is used in the experiment, is
shown in Fig. \ref{fig_EML_TransferCurve}, where we represent the EML output power as a function of the bias voltage.
The operation bias point is generally chosen approximately in the inflection point of the transfer curve to utilize
homogeneously the upper and lower portion of the transfer curve. For this particular scenario, the considered EML would
be operated at the bias voltage of $-3.1$~V.

The EML transfer function can be expressed
as~\cite{EML_DML_MZM_Zhang:18}
\begin{align}
	H_{EML} = |\sqrt{1+\beta^2}~\cos(\theta + \arctan~\beta)|.  \label{eq_EML_transferFunction}
\end{align} 
Here $\beta$ is the chirp parameter of the EML, which describes the relationship between intensity modulation and phase
modulation. $\beta$ varies with the bias voltage: A larger negative value of $\beta$ is expected for larger negative
bias voltages and vice-versa~\cite{EML_DML_MZM_Zhang:18}. The EML transfer curve has two distinctive features:
\begin{itemize} 
	\item It shows a slow transition from the linear to the nonlinear region.
	\item Its chirp ($\beta$) varies with the bias voltage.
\end{itemize}

Fig. \ref{fig_EML_TransferCurve} illustrates the modulation of a sine wave. The EML is biased approximately in the inflection point of the linear region. By regulating
the driving voltage we can control the ER and the used nonlinear region of the EML transfer curve.
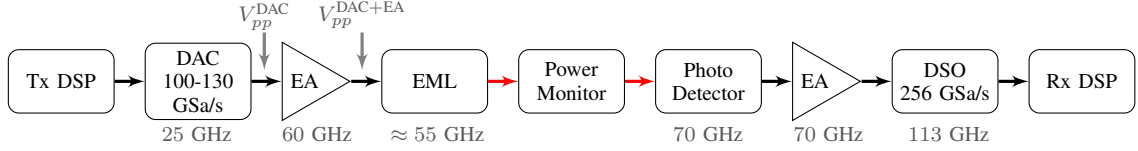
\begin{figure*}[t] 
	\footnotesize
	\centering
	\tikzset{
	label/.style = {draw=none, fill=none, rectangle, minimum height=2em, minimum width=2em},
	block/.style = {draw, fill=white, rectangle, minimum height=3em, minimum width=5em,rounded corners},
	block1/.style = {draw, fill=none, rectangle, minimum height=8em, minimum width=12em,rounded corners},
	block2/.style = {draw, fill=white,color=white, rectangle, minimum height=0.7em, minimum width=11em,rounded corners},
	amplifiersrec/.style = {draw, fill=white, rectangle, minimum height=1em, minimum width=5em,rounded corners},
	blockrx/.style = {draw, fill=white, rectangle, minimum height=3em, minimum width=6em,rounded corners},
	amplifiers/.style = {draw, fill=white, regular polygon, regular polygon sides=3,minimum size = 1cm},
	EDFA/.style = {draw, fill=white, regular polygon, regular polygon sides=3,minimum size=1.2cm},
	tmp/.style  = {coordinate}, 
	sum/.style= {draw, fill=none, circle, node distance=2cm,color=red,thick,minimum size=27pt},
	voa/.style= {draw, fill=none, circle, node distance=2cm,color=red,thick,minimum size=20pt},
	input/.style = {coordinate},
	output/.style= {coordinate},
	pinstyle/.style = {pin edge={to-,thin,black}}
}

	\begin{tikzpicture}[auto, node distance=0.3cm,>=latex']
		\node [block] (TX_DSP) {Tx DSP};
		\node [block, right=0.4cm of TX_DSP,align = center] (DAC) {DAC \\ 100-130 \\GSa/s};
		\node [amplifiers, right=0.4cm of DAC,align = center,rotate=0,shape border rotate=-90, scale=1.2] (EA1) {};
		\node [block, right=0.4cm of EA1,align = center] (EML) {EML};
		
		\node [block, right=0.4cm of EML,align = center] (PowerMonitor) {Power\\ Monitor};
		\node [block, right=0.4cm of PowerMonitor,align = center] (PhotoDetector) {Photo\\ Detector};
		\node [amplifiers, right=0.4cm of PhotoDetector,align = center,rotate=0,shape border rotate=-90, scale=1.2] (EA2) {};
		\node [block, right=0.4cm of EA2,align = center] (DSO) {DSO \\ 256 GSa/s};
		\node [block, right=0.4cm of DSO,align = center] (RxDSP) {Rx DSP};
		
		\draw[white,anchor=base] (0,-0.7) -- (13,-0.5)
		node[black!65,shift={(0,-0.8)}] at (1.8,0) {$25$ GHz}
		node[black!65,shift={(0,+0.8)}] at (2.7,0) {$V_{pp}^{\textnormal{DAC}}$}
		node[black!65,shift={(0,-0.8)}] at (3.4,0) {$60$ GHz}
		node[black!65,shift={(0,+0.8)}] at (4.0,0) {$V_{pp}^{\textnormal{DAC$+$EA}}$}
		node[black!65,shift={(0,-0.8)}] at (5,0) {$\approx55$ GHz}
		node[black!65,shift={(0,-0.8)}] at (8.6,0) {$70$ GHz}
		node[black!65,shift={(0,-0.8)}] at (10.2,0) {$70$ GHz}
		node[black!65,shift={(0,-0.8)}] at (11.8,0) {$113$ GHz};
		
		\coordinate (a) at (2.7,0.65);
		\coordinate (b) at (2.7,0.15);
		\draw[->,very thick,color=gray] (a) -- (b);
		
		\coordinate (c) at (4.0,0.65);
		\coordinate (d) at (4.0,0.15);
		\draw[->,very thick,color=gray] (c) -- (d);

		\node [black] at (EA1.center) {EA};
		\node [black] at (EA2.center) {EA};
		
		\draw [->,very thick] (TX_DSP) -- (DAC);
		\draw [->,very thick] (DAC) -- (EA1);
		\draw [->,very thick] (EA1) -- (EML);
		\draw [->,very thick,color=red] (EML) -- (PowerMonitor);
		\draw [->,very thick,color=red] (PowerMonitor) -- (PhotoDetector);
		\draw [->,very thick] (PhotoDetector) -- (EA2);
		\draw [->,very thick] (EA2) -- (DSO);
		\draw [->,very thick] (DSO) -- (RxDSP);
		
	\end{tikzpicture}
	\vspace{3mm}
	\caption{Schematic of the IM/DD transmission system used for the experimental investigations. DAC: digital-to-analog
		converter, EA: electrical amplifier, EML: electro-absorption modulated laser, DSO: digital storage
		oscilloscope.}
	\label{fig_blockdiagram_setup}
\end{figure*}
The optimum driving voltage highly depends on the modulation format and the symbol occurrence probability. As shown in
Fig. \ref{fig_PS_SymbolDistribution}, with the cup-shaped MB distribution, the outer levels have higher occurrence
probability and, therefore, are affected by distortion which leads to higher noise variance. For the cup distribution, the optimum $V_{pp}$ is generally lower in comparison to both cap-shaped and uniform
PAM-8. The cap distribution has the lowest symbol occurrence on the outer levels and tolerates a high $V_{pp}$ without
suffering from nonlinear distortions. Therefore, the cap-shaped distribution allows for ER enhancement, which leads to
better separation between neighboring levels and improves the bit error ratio (BER) performance. For uniform PAM-8, the optimum $V_{pp}$
is in between the optimal values for cap and cup-shaped distributions.

Chirp dependence on bias voltage also plays an important role when fiber transmission is performed for the comparison
between uniform and PS PAM-8. As explained, higher negative and lower negative bias voltages result in high negative and
positive $\beta$ values, respectively. The variation of the $\beta$ values results in nonlinear distortions. For the cap
variant, most of the symbols are transmitted around the zero level, where $|\beta|$ is smaller, which reduces the impact
of nonlinear distortion. Conversely, the cup variant is more susceptible to nonlinear distortion since most symbols are
modulated on the outer levels where $|\beta|$ is higher. Uniform PAM-8 is expected to have an intermediate behavior.

In summary, for an EML, the cap-shaped distribution exhibits two distinct beneficial features:
\begin{itemize} 
	\item $V_{pp}$ $\uparrow$ $\Rightarrow$ ER $\uparrow$ $\Rightarrow$  separation between neighboring
	levels~$\uparrow$ $\Rightarrow$ BER $\downarrow$.
	\item Most symbol are modulated around the zero level, where $|\beta|$ is not significant $\Rightarrow$ nonlinear
	distortion $\downarrow$ (only valid when transmission is performed over dispersive fiber link).
\end{itemize}
\section{Experimental Setup and Transceiver DSP}\label{sec_experimentalSetup}

In this section, the experimental setup and the digital signal processing (DSP) used for experimental evaluation are
introduced.

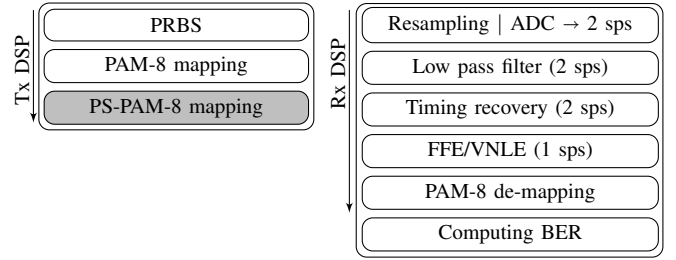
\begin{figure}[t] 
	\footnotesize
	\centering
	\tikzset{
	label/.style = {draw, fill=none, rectangle, minimum height=2em, minimum width=3em},
	border1/.style = {draw=black, rectangle, minimum height=6em, minimum width=13em,rounded corners},
	border2/.style = {draw=black, rectangle, minimum height=12em, minimum width=14.5em,rounded corners},
	blockrx/.style = {draw, fill=white, rectangle, minimum height=1.5em, minimum width=3.5em,rounded corners},
	blocktx/.style = {draw, fill=white, rectangle, minimum height=1.5em, minimum width=12.5em,rounded corners},
	block/.style = {draw, fill=white, rectangle, minimum height=1.5em, minimum width=14em,rounded corners},
	blocktx2/.style = {draw, fill=gray!50, rectangle, minimum height=1.5em, minimum width=12.5em,rounded corners},
	tmp/.style  = {coordinate}, 
	sum/.style= {draw, fill=white, circle, node distance=1cm},
	input/.style = {coordinate},
	output/.style= {coordinate},
	pinstyle/.style = {pin edge={to-,thin,black}
	}
}

	\begin{tikzpicture}[auto, node distance=0.3cm,>=latex']

		\node [label, draw=white,rotate=90] at (-2.35,-0.6)  (channeloutput) {Tx DSP};
		\node [blocktx] at (-0.3,0)   (PRBS) {PRBS};
		\node [blocktx,below=0.1cm of PRBS]  (Mapping) {PAM-8 mapping};
		\node [blocktx2,below=0.1cm of Mapping]  (PS-PAM) {PS-PAM-8 mapping};
		\node [border1] at (-0.3,-0.55) () {};
		\draw [->] (-2.2,0.2) -- (-2.2,-1.3);

		\node [label, draw=white,rotate=90] at (1.85,-0.65)  (channeloutput) {Rx DSP};
		\node [block,right=0.7cm of PRBS]  (Resampling) {Resampling $|$ ADC $\shortrightarrow$ 2~sps};
		\node [block,below=0.1cm of Resampling]  (LPF) {Low pass filter (2~sps)};	
		\node [block,below=0.1cm of LPF]  (Timing Recovery) {Timing recovery (2~sps)};			
		\node [block,below=0.1cm of Timing Recovery]  (VNLE) {FFE/VNLE (1~sps)};
		\node [block,below=0.1cm of VNLE]  (demapping) {PAM-8 de-mapping};
		\node [block,below=0.1cm of demapping]  (computingBER) {Computing BER};
		\node [border2] at (4.13,-1.4) () {};
		
		\draw [->] (2.0,0.2)  -- (2.0,-2.5) ;	
		
	\end{tikzpicture}
	\caption{Block diagram of the offline DSP stack.}
	\label{fig_dsp_stack}
\end{figure}

Fig. \ref{fig_blockdiagram_setup} depicts the optically unamplified IM/DD setup, which is utilized for
experimental evaluations. Below each OE component, its $3$-dB BW is indicated. Both the transmitter (Tx) and receiver (Rx)
side use offline DSP. The block diagrams of the Tx and Rx DSP stacks are shown in Fig. \ref{fig_dsp_stack}. For the
transmitter side, a pseudo-random binary sequence (PRBS) is generated and mapped on the uniform PAM-8 levels.
Alternatively, for PS-PAM-8, pseudo-random symbols are drawn from an MB distribution with the desired entropy and GO.
For uniform PAM-8, the symbol rate is varied from $100$ GBd to $130$ GBd in $5$~GBd steps. For both cap-shaped (Fig. \ref{fig_PS_SymbolDistribution} (a)) and
cup-shaped (Fig. \ref{fig_PS_SymbolDistribution} (b)) distributions, a PS OH (Eq. \ref{eq:PS_OH}) of $\approx8.17\%$ is kept constant throughout the experiments. For the cap-shaped (Fig. \ref{fig_PS_SymbolDistribution} (c)) variant, GOs of  $2.0, 3.5~\textnormal{and}~5.0$  are used and symbol rates of $110$~GBd, $120$~GBd and $130$~GBd are transmitted. In case of the cup-shaped
MB distribution, only a GO of $2.0$ at $110$~GBd is evaluated.

The mapped symbols are quantized and fed to the digital-to-analog converter (DAC), which is operated at $1$ sample per symbol (sps) and has a $3$-dB
bandwidth (BW) of $25$~GHz ($6$-dB BW $\approx$ $40$~GHz). By changing the frequency of its clock source, the sampling rate of
the DAC is varied from $100$ to $130$~GSa/s.Additionally, we can also vary the voltage swing of the DAC, which we define as $V_{pp}^{\textnormal{DAC}}$ in mV unit. After the DAC, an electrical amplifier (EA) of $60$~GHz BW with $22$~dB gain is
used. The resultant amplified electrical output voltage is controlled with the $V_{pp}^{\textnormal{DAC}}$ and the output voltage of the DAC and EA, which is also the driving voltage of the modulator, is defined as $V_{pp}^{\textnormal{DAC$+$EA}}$. In Practice, it is difficult to measure the effective $V_{pp}^{\textnormal{DAC$+$EA}}$. Since the estimation of $V_{pp}^{\textnormal{DAC$+$EA}}$ is dubious, while $V_{pp}^{\textnormal{DAC}}$ is known and regulated by us during the experiments, throughout the paper, we report our results w.r.t. $V_{pp}^{\textnormal{DAC}}$. The amplified signal with $V_{pp}^{\textnormal{DAC$+$EA}}$ is then modulated with an O-band EML ($3$-dB BW $\approx$ $55$~GHz), whose transfer curve is
presented in Fig. \ref{fig_EML_TransferCurve}. During the experiments, a current of $66$~mA is supplied to the integrated
laser diode of the EML and a temperature of $\approx25$ \textdegree~C is maintained for the packaged EML. The EML bias voltage
is set to $-3.1$~V, which leads to an EML output power of $\approx2.8$ mW ($4.5$ dBm). The driving voltage is regulated via
the analog gain of the DAC. At high values of $V_{pp}$, the nonlinear compression of the modulator can affect the
average output power by up to $\pm0.1$~dBm. Since the variation is not significant, for simplicity, the EML output
power is considered $\approx4.5$~dBm independent of the driving voltage.  An optional inline power monitor is inserted  to measure the output power of the EML, which is also the received optical power (ROP) impinging the $70$-GHz PIN PD. 

At the receiver, the optical signal is detected and converted to an electrical signal by the PIN PD.  Since a proper
trans-impedance amplifier was unavailable during the experiments, the electrical signal was amplified with a
$70$~GHz $11$~dB fixed gain EA. The resulting electrical signal is captured with a digital storage oscilloscope (DSO), which
operates at $256$~GSa/s and has a $3$-dB BW of 113~GHz.

For offline processing, $4$~million samples are used. The captured samples are first resampled to $2$ sps according to the
transmission symbol rate. Afterwards, the received signal is normalized by removing the DC components in the signal.
Later a low-pass filter designed as finite impulse response filter with filter order $30$ is applied to cut the
out-of-band noise and prevent aliasing. To estimate and compensate jitter and sampling frequency offset, timing recovery
is performed~\cite{Nebojsa_ABSPD_OFC20}. Either linear feed-forward equalization (FFE) or Volterra nonlinear
equalization (VNLE) is performed on every other symbol ($1$~sps). In both cases, $301$ linear equalizer taps are used to
minimize the effect of reflections from the discrete component setup, long copper cables and connectors. For the DSP
configuration with VNLE, besides the linear filter taps, additionally $2$\textsuperscript{nd} and $3$\textsuperscript{rd}
order Volterra kernels are used with memory lengths of $7$ and $9$ symbols to compensate for nonlinearity introduced by DAC,
EA, EML and square-law detection, respectively. After equalization (FFE/VNLE), the PAM-8 symbols are de-mapped assuming
Gray mapping and the BER is computed.
 
\section{Experimental Result and Discussion} \label{sec_experimentalresult_discussion}

An experimental comparison of uniform PAM-8 and PS PAM-8 with different GOs is presented in this section. The
adopted performance metric is pre-FEC BER. Later we interpret the BER result as achievable information rate (AIR) and operational net bit rate (ONBR).

\subsection{Comparison between Uniform and PS PAM-8} \label{subsec_uniform_cap_cup}

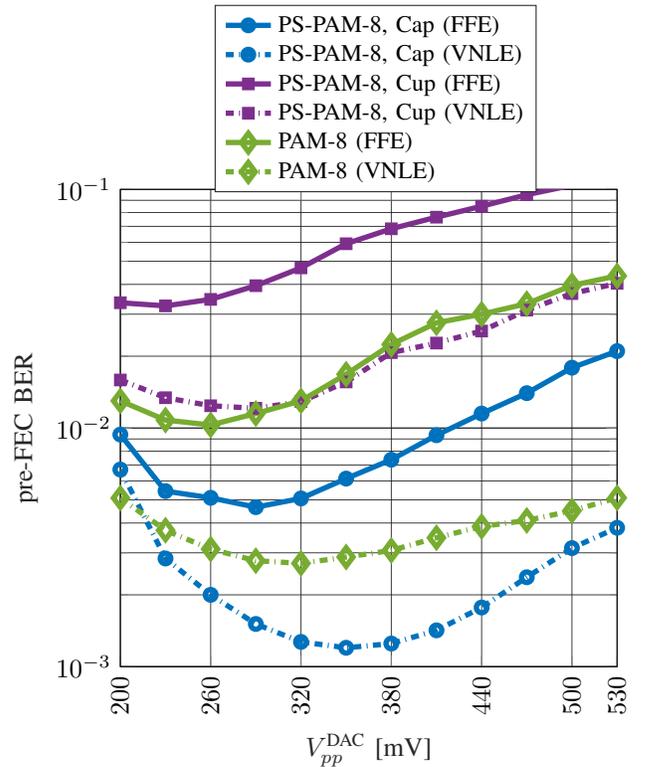
\begin{figure}[t]  
	\centering
%
%
\definecolor{mycolor1}{rgb}{0.00000,0.44700,0.74100}%
\definecolor{mycolor2}{rgb}{0.49400,0.18400,0.55600}%
\definecolor{mycolor3}{rgb}{0.46600,0.67400,0.18800}%
\begin{tikzpicture}

\begin{axis}[%
width=2.6in,
height=2.5in,
at={(0.758in,0.676in)},
scale only axis,
xmin=200,
xmax=530,
xtick={200,260,320,380,440,500,530},
xticklabel style={rotate=90},
xlabel style={font=\color{white!15!black}},
xlabel={$V_{pp}^{\textnormal{DAC}}$ [mV]},
ymode=log,
ymin=0.001,
ymax=0.1,
yminorticks=true,
ylabel style={font=\color{white!15!black}},
ylabel={pre-FEC BER},
axis background/.style={fill=white},
xmajorgrids,
ymajorgrids,
yminorgrids,
grid style={opacity=0.85},
minor grid style={opacity=0.85},
grid style={line width=.1pt, draw=black!90},
legend style={at={(0.838,0.994)},{nodes={scale=0.9, transform shape}}, anchor=south east, legend cell align=left, align=left, draw=white!15!black}
]
\addplot [color=mycolor1, line width=2.0pt, mark size=2.0pt, mark=o, mark options={solid, mycolor1}]
  table[row sep=crcr]{%
200	0.00939\\
230	0.00544\\
260	0.0051\\
290	0.00466\\
320	0.00507\\
350	0.00614\\
380	0.00737\\
410	0.00933\\
440	0.0115\\
470	0.014\\
500	0.0179\\
530	0.021\\
};
\addlegendentry{PS-PAM-8, Cap (FFE)}

\addplot [color=mycolor1, dashdotted, line width=2.0pt, mark size=2.0pt, mark=o, mark options={solid, mycolor1}]
  table[row sep=crcr]{%
200	0.0067\\
230	0.00284\\
260	0.002\\
290	0.00151\\
320	0.00127\\
350	0.0012\\
380	0.00125\\
410	0.00142\\
440	0.00177\\
470	0.00237\\
500	0.00314\\
530	0.00382\\
};
\addlegendentry{PS-PAM-8, Cap (VNLE)}

\addplot [color=mycolor2, line width=2.0pt, mark size=1.4pt, mark=square, mark options={solid, mycolor2}]
  table[row sep=crcr]{%
200	0.0335\\
230	0.0325\\
260	0.0346\\
290	0.0395\\
320	0.0469\\
350	0.0592\\
380	0.0684\\
410	0.0765\\
440	0.085\\
470	0.095\\
500	0.104\\
530	0.112\\
};
\addlegendentry{PS-PAM-8, Cup (FFE)}

\addplot [color=mycolor2, dashdotted, line width=2.0pt, mark size=1.4pt, mark=square, mark options={solid, mycolor2}]
  table[row sep=crcr]{%
200	0.0159\\
230	0.0134\\
260	0.0124\\
290	0.0121\\
320	0.0129\\
350	0.0156\\
380	0.0207\\
410	0.0227\\
440	0.0255\\
470	0.0312\\
500	0.0366\\
530	0.0404\\
};
\addlegendentry{PS-PAM-8, Cup (VNLE)}

\addplot [color=mycolor3, line width=2.0pt, mark size=3.5pt, mark=diamond, mark options={solid, mycolor3}]
  table[row sep=crcr]{%
200	0.013\\
230	0.0108\\
260	0.0103\\
290	0.0115\\
320	0.013\\
350	0.0168\\
380	0.0224\\
410	0.0276\\
440	0.03\\
470	0.0332\\
500	0.0396\\
530	0.0434\\
};
\addlegendentry{PAM-8 (FFE)}

\addplot [color=mycolor3, dashdotted, line width=2.0pt, mark size=3.5pt, mark=diamond, mark options={solid, mycolor3}]
  table[row sep=crcr]{%
200	0.00509\\
230	0.00373\\
260	0.00311\\
290	0.00278\\
320	0.00271\\
350	0.00288\\
380	0.00307\\
410	0.00347\\
440	0.00387\\
470	0.0041\\
500	0.0045\\
530	0.0051\\
};
\addlegendentry{PAM-8 (VNLE)}

\end{axis}

\end{tikzpicture}%
	\caption{Driving voltage optimization for uniform and PS \hbox{PAM-8} with cap and cup-shaped MB distributions (GO $2.0$).}
	\label{fig_BER_cap_or_cup}
\end{figure}

The optimization of the driving voltage has been discussed in Section \ref{sec_DrivingVoltage_optimization}. Now
experimental results for uniform \hbox{PAM-8}, and cap and cup-shaped PS PAM-8 with GO $2.0$ are presented in \hbox{Fig. \ref{fig_BER_cap_or_cup}} in the form of pre-FEC BER as a function of $V_{pp}^{\textnormal{DAC}}$. The plot compares $100$ GBd uniform PAM-8 with cap and cup-shaped
variants (with GO $2.0$) with a symbol rate of 110 GBd and PS OH of $\approx8.17\%$. Considering an HD-FEC with 7\% OH (BER threshold of $4.6\times 10^{-3}$ \cite{MichaelScholten_HD_FEC_ECOC_workshop}), the
three formats achieve a similar net bit rate (NBR) of $\approx280$~Gb/s (the NBR for uniform and PS PAM-8 is
$280.37$~Gb/s and $283.48$~Gb/s, respectively).

Fig. \ref{fig_BER_cap_or_cup} presents results obtained with linear (solid lines) and nonlinear (dashed-dotted lines)
equalization. For the DSP with FFE only, uniform PAM-8 achieves the lowest BER $1.3\times 10^{-2}$ at $260$~mV. While the cup-shaped (purple color, square markers) MB variant performs the worst and corresponds to an optimum $V_{pp}^{\textnormal{DAC}}$ of $230$~mV. Conversely, the cap-shaped (blue color, circle markers) MB variant performs the best with BER improved by more than $7$ times compared to uniform PAM-8 at the optimum $V_{pp}^{\textnormal{DAC}}$ of $290$ mV. As expected, uniform PAM-8
(green color, diamond markers) performs at an intermediate level with the optimum $V_{pp}^{\textnormal{DAC}}$ at exactly halfway between
the two PS formats. In the presence of VNLE, the performance of all compared schemes is improved. Also, the optimum
driving voltage increases significantly in the presence of VNLE. A significant BER improvement is achieved in case of
uniform PAM-8 (more than $7$ times), while for the cup and cap-shaped distributions, the improvement factor reaches $3$ times at most.
\begin{figure}[t]  
	\centering
	\input{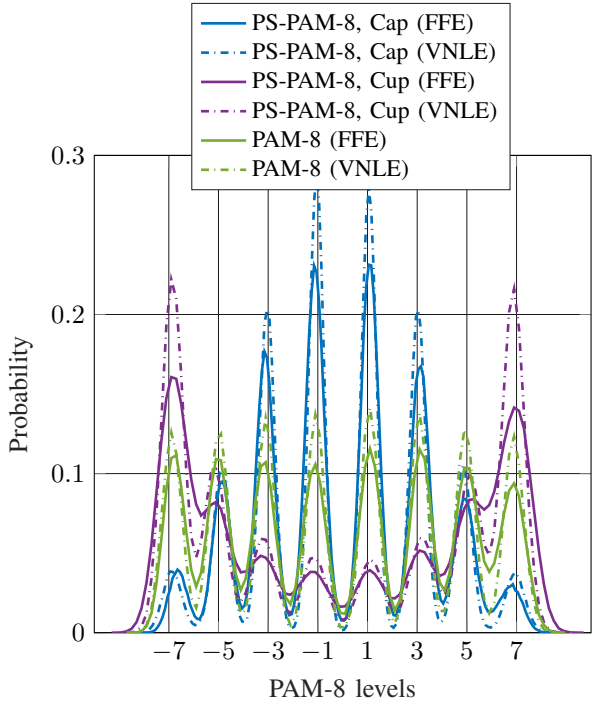}
	\caption{Histograms of uniform PAM-8 and PS PAM-8 with linear and nonlinear equalization.}
	\label{fig_histogram_cap_or_cup}
\end{figure}

Histograms of uniform and PS PAM-8 at their respective optimum $V_{pp}^{\textnormal{DAC}}$ after equalization are presented in Fig.
\ref{fig_histogram_cap_or_cup}. Line color and style are kept the same as in Fig. \ref{fig_BER_cap_or_cup}. For the
cap-shaped PS variant,  the linear FFE results in a very distinct level separation, while VNLE compensates the minor
nonlinear distortions. For the cup-shaped MB distribution, the VNLE is essential to correct the significant nonlinear
distortions. In case of uniform PAM-8, all levels exhibit a reasonable separation in the presence of linear equalization
and, at the same time, largely benefit from VNLE.

In summary, uniform PAM-8 is highly susceptible to nonlinear distortion, and cap-shaped PS-PAM-8 with GO $2.0$ performs the best, since it allows the highest driving voltage and exhibits the highest nonlinear tolerance. Therefore, we conclude that in the
considered scenario, probabilistic constellation shaping with cap-shaped distributions results in nonlinear shaping gain
over uniform PAM-8 at the cost of increased symbol rate.

Additionally, while sweeping the driving voltage during $V_{pp}$ optimization, we are deliberately operating both in the linear and nonlinear region of the EML. The left part of the curves in Fig. 5 corresponds to the linear region. If we drive the system in the linear region, we lose performance because of the lower transmit power. From the trend in Fig. 5, it can be extrapolated that the performance gain of PS-PAM-8 with cap shape over uniform PAM-8 will vanish at lower driving voltage.

\vspace{-4mm}

\subsection{Comparison among Different Gaussian Orders} \label{subsec_cap_differentGaussianOrders}

We have introduced the concept of GO in Eq. \ref{eq_Cap_Gaussian_order}. Since the cap-shaped MB distribution
undoubtedly performs better than the cup-shaped variant, we concentrate our investigation on different GOs
on cap-shaped distributions only.
\begin{figure}[t]
	\centering
%
%
\definecolor{mycolor1}{rgb}{0.00000,0.44700,0.74100}%
\definecolor{mycolor2}{rgb}{0.85000,0.32500,0.09800}%
\definecolor{mycolor3}{rgb}{0.92900,0.69400,0.12500}%
\begin{tikzpicture}

\begin{axis}[%
width=2.6in,
height=2.5in,
at={(0.555in,0.53in)},
scale only axis,
xmin=200,
xmax=530,
xtick={200,260,320,380,440,500,530},
xticklabel style={rotate=90},
xlabel style={font=\color{white!15!black}},
xlabel={$V_{pp}^{\textnormal{DAC}}$ [mV]},
ymode=log,
ymin=0.001,
ymax=0.03,
yminorticks=true,
ylabel style={font=\color{white!15!black}},
ylabel={pre-FEC BER},
axis background/.style={fill=white},
xmajorgrids,
ymajorgrids,
yminorgrids,
grid style={opacity=0.85},
minor grid style={opacity=0.85},
grid style={line width=.1pt, draw=black!90},
legend style={at={(0.72,0.7)}, {nodes={scale=0.9, transform shape}},anchor=south east, legend cell align=left, align=left, draw=white!15!black}
]
\addplot [color=mycolor1, line width=2.0pt, mark size=2.0pt, mark=o, mark options={solid, mycolor1}]
  table[row sep=crcr]{%
200	0.00939\\
230	0.00544\\
260	0.0051\\
290	0.0046\\
320	0.00507\\
350	0.00614\\
380	0.00737\\
410	0.00933\\
440	0.0115\\
470	0.014\\
500	0.0179\\
530	0.021\\
};
\addlegendentry{Gaussian order: 2.0 (FFE)}

\addplot [color=mycolor1, dashdotted, line width=2.0pt, mark size=2.0pt, mark=o, mark options={solid, mycolor1}]
  table[row sep=crcr]{%
200	0.0067\\
230	0.00284\\
260	0.002\\
290	0.00151\\
320	0.00127\\
350	0.0012\\
380	0.00125\\
410	0.00142\\
440	0.00177\\
470	0.00237\\
500	0.00314\\
530	0.00382\\
};
\addlegendentry{Gaussian order: 2.0 (VNLE)}

\addplot [color=mycolor2, line width=2.0pt, mark size=1.4pt, mark=square, mark options={solid, mycolor2}]
  table[row sep=crcr]{%
200	0.00954\\
230	0.0064\\
260	0.00505\\
290	0.00446\\
320	0.00426\\
350	0.0050\\
380	0.00603\\
410	0.0075\\
440	0.009\\
470	0.0115\\
500	0.0156\\
530	0.019\\
};
\addlegendentry{Gaussian order: 3.5 (FFE)}

\addplot [color=mycolor2, dashdotted, line width=2.0pt, mark size=1.4pt, mark=square, mark options={solid, mycolor2}]
  table[row sep=crcr]{%
200	0.00657\\
230	0.0028\\
260	0.00195\\
290	0.00145\\
320	0.0012\\
350	0.0011\\
380	0.0011\\
410	0.00113\\
440	0.00147\\
470	0.00185\\
500	0.00251\\
530	0.00317\\
};
\addlegendentry{Gaussian order: 3.5 (VNLE)}

\addplot [color=mycolor3, line width=2.0pt, mark size=1.3pt, mark=triangle, mark options={solid, mycolor3}]
  table[row sep=crcr]{%
200	0.00836\\
230	0.0064\\
260	0.00505\\
290	0.00436\\
320	0.00420\\
350	0.00406\\
380	0.00503\\
410	0.0066\\
440	0.008\\
470	0.0105\\
500	0.0136\\
530	0.0181\\
};
\addlegendentry{Gaussian order: 5.0 (FFE)}

\addplot [color=mycolor3, dashdotted, line width=2.0pt, mark size=1.3pt, mark=triangle, mark options={solid, mycolor3}]
  table[row sep=crcr]{%
200	0.00546\\
230	0.00276\\
260	0.00186\\
290	0.0014\\
320	0.00118\\
350	0.00108\\
380	0.00103\\
410	0.00108\\
440	0.0014\\
470	0.00189\\
500	0.00231\\
530	0.00299\\
};
\addlegendentry{Gaussian order: 5.0 (VNLE)}

\end{axis}

\begin{axis}[%
width=2.952in,
height=2.756in,
at={(0in,0in)},
scale only axis,
xmin=0,
xmax=1,
ymin=0,
ymax=1,
axis line style={draw=none},
ticks=none,
axis x line*=bottom,
axis y line*=left
]
\node[fill=white, below right, align=left, draw=black,rounded corners]
at (rel axis cs:0.383,0.483) {PS-PAM-8 $110$ GBd\\Cap, PS OH: $8.17\%$};
\end{axis}
\end{tikzpicture}%
	\caption{Driving voltage optimization for cap-shaped distributions of different GOs.}
	\label{fig_BER_differentGaussianOrders}
\end{figure}
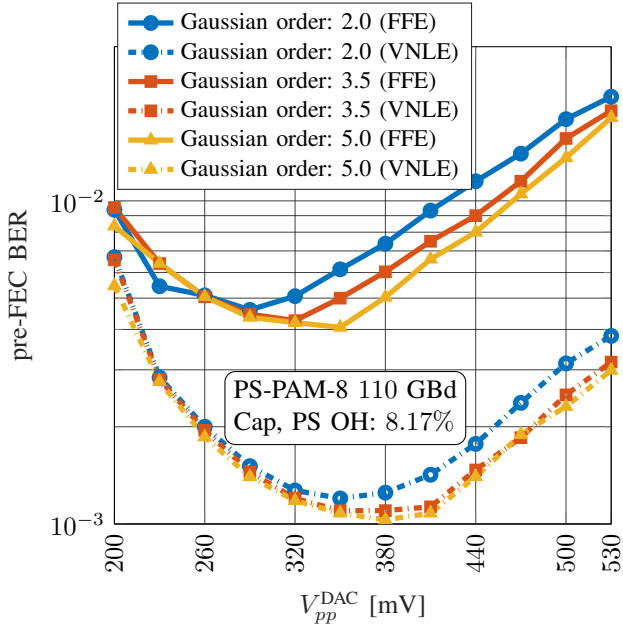
\begin{figure}[t]  
	\centering
	\input{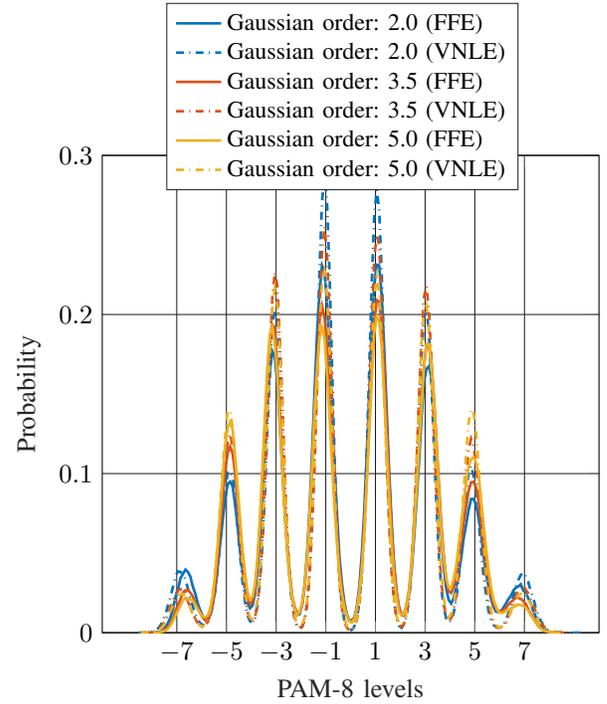}
	\caption{Signal histograms for cap-shaped distributions of different Gaussian orders with and without nonlinear
		equalization.}
	\label{fig_Histogram_differentGaussianOrders}
\end{figure}

Fig. \ref{fig_BER_differentGaussianOrders} shows the experimental results for cap-shaped \mbox{PS-PAM-8} at 110 GBd with PS OH
$\approx8.17\%$ and GO $2.0, 3.5~\textnormal{and}~5.0$. Similar to the previous investigations, results are presented here in the case of
linear and nonlinear equalization. In the case of equalization with linear FFE, for GO 5.0, the optimum $V_{pp}$ is
60~mV higher than for GO $2.0$ and $30$~mV higher than for GO $3.5$.  As explained in Section \ref{sec_probabilistic_shaping},
the optimum $V_{pp}$ depends on the probability of the symbols mapped on the outer levels. For higher GOs the
distribution becomes more compact and a higher $V_{pp}$ can be tolerated. However, in our experiment, the BER
improvement due to higher-order GOs is only minor. With VNLE, the optimum driving voltage is increased by $60$~mV for all
considered GOs and the BER improvement of GO $5.0$ compared to GO $2.0$ is marginal. To understand the phenomena, signal
histograms for different GOs are presented at their optimum $V_{pp}$ in \hbox{Fig.~\ref{fig_Histogram_differentGaussianOrders}}. In the presence of VNLE, the BER is improved up to $3$ times, but the
different GOs perform similarly, because, after equalizing the nonlinearity, the main residual impairment is the BW
limitation of OE components. Therefore, we can see that going beyond GO $5.0$ does not bring significant benefits. Please note that this is different from the scenario considered in \cite{Hossain_PS_PAM_ECOC22}, where we compared higher-order GOs in the case of less severe bandwidth limitation. In~\cite{Hossain_PS_PAM_ECOC22} GO $5.0$, in term of ROP, performs on average $0.5$ to $0.8$ dB better compared to GO $2.0$. In this scenario~\cite{Hossain_PS_PAM_ECOC22}, we can see that higher GO brings always gain and going beyond GO $5.0$ may improve performance further.

Since we assume an HD-FEC, we consider the commonly used pre-FEC BER as performance metric. Note that the AIR for HD is a monotonic function of the BER (computing AIR is described in Sec. V$-$C). Consequently, optimizing w.r.t. BER or HD-AIR leads to the same optimization outcome.

\subsection{BER, AIR and ONBR vs Symbol Rate} \label{subsec_BER_AIR_vs_SymbolRate}

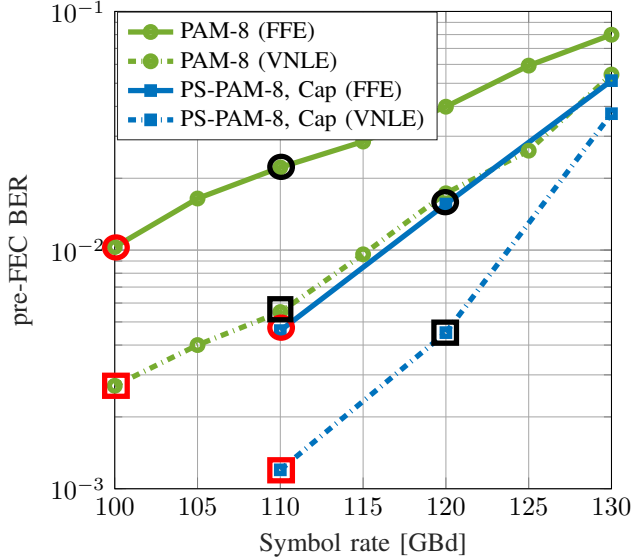
\begin{figure}[t]  
	\centering
%
%
\definecolor{mycolor1}{rgb}{0.46600,0.67400,0.18800}%
\definecolor{mycolor2}{rgb}{0.00000,0.44700,0.74100}%
\definecolor{mycolor3}{rgb}{1.00000,0.41176,0.16078}%
\begin{tikzpicture}

\begin{axis}[%
width=2.6in,
height=2.5in,
at={(0.758in,0.562in)},
scale only axis,
xmin=100,
xmax=130,
xlabel style={font=\color{white!15!black}},
xlabel={Symbol rate [GBd]},
ymode=log,
ymin=0.001,
ymax=0.1,
yminorticks=true,
ylabel style={font=\color{white!15!black}},
ylabel={pre-FEC BER},
axis background/.style={fill=white},
xmajorgrids,
ymajorgrids,
yminorgrids,
grid style={line width=.1pt, draw=gray!70},
legend style={at={(0.65,0.73)}, {nodes={scale=0.9, transform shape}},anchor=south east, legend cell align=left, align=left, draw=white!15!black}
]
\addplot [color=mycolor1, line width=2.0pt, mark size=2.0pt, mark=o, mark options={solid, mycolor1}]
  table[row sep=crcr]{%
100	0.0103\\
105	0.01645\\
110	0.02221\\
115	0.0285\\
120	0.0399\\
125	0.0593\\
130	0.0799\\
};
\addlegendentry{PAM-8 (FFE)}

\addplot [color=mycolor1, dashdotted, line width=2.0pt, mark size=2.0pt, mark=o, mark options={solid, mycolor1}]
  table[row sep=crcr]{%
100	0.0027\\
105	0.0040\\
110	0.0055\\
115	0.0096\\
120	0.0173\\
125	0.0261\\
130	0.0544\\
};
\addlegendentry{PAM-8 (VNLE)}

\addplot [color=mycolor2, line width=2.0pt, mark size=1.4pt, mark=square, mark options={solid, mycolor2}]
  table[row sep=crcr]{%
110	0.0046\\
120	0.0156\\
130	0.0513\\
};
\addlegendentry{PS-PAM-8, Cap (FFE)}

\addplot [color=mycolor2, dashdotted, line width=2.0pt, mark size=1.4pt, mark=square, mark options={solid, mycolor2}]
  table[row sep=crcr]{%
110	0.0012\\
120	0.00451\\
130	0.0373\\
};
\addlegendentry{PS-PAM-8, Cap (VNLE)}

\end{axis}

\begin{axis}[%
width=2.952in,
height=2.756in,
at={(0in,0in)},
scale only axis,
xmin=0,
xmax=1,
ymin=0,
ymax=1,
axis line style={draw=none},
ticks=none,
axis x line*=bottom,
axis y line*=left
]
\draw [red, line width=2.0pt] (axis cs:0.258893,0.66248) ellipse [x radius=0.02, y radius=0.02];
\draw [red, line width=2.0pt] (axis cs:0.55175,0.51) ellipse [x radius=0.02, y radius=0.02];
\draw [black, line width=2.0pt] (axis cs:0.55175,0.81565952) ellipse [x radius=0.02, y radius=0.02];
\draw [black, line width=2.0pt] (axis cs:0.842679,0.748524) ellipse [x radius=0.02, y radius=0.02];

\draw [red, thick,line width=2.0pt] (axis cs:0.238893,0.379571) rectangle (axis cs:0.278893,0.419571);
\draw [red, thick,line width=2.0pt] (axis cs:0.53175,0.21895) rectangle (axis cs:0.57175,0.25895);

\draw [black, thick,line width=2.0pt] (axis cs:0.53175,0.525) rectangle (axis cs:0.568893,0.565);
\draw [black, thick,line width=2.0pt] (axis cs:0.822679,0.4805242) rectangle (axis cs:0.862679,0.5205242);

\end{axis}
\end{tikzpicture}%
	\caption{Experimental comparison of uniform and cap-shaped PS PAM-8.}
	\label{fig_BERvsSymbolRate}
\end{figure}

In this section we will concentrate our investigation on uniform PAM-8 and cap-shaped PS-PAM-8 with GO $2.0$. Experimental results in terms of pre-FEC BER as a function of the symbol rate are presented in Fig.
\ref{fig_BERvsSymbolRate}. Considering an HD-FEC with $7\%$ OH (BER threshold of $4.6\times 10^{-3}$)~\cite{MichaelScholten_HD_FEC_ECOC_workshop}, uniform PAM-8 at $100$~GBd (red circle (FFE), square (VNLE)) and $110$~GBd (red circle FFE, square (VNLE)) is approximately comparable with \mbox{PS-PAM-8}  with PS OH of $\approx8.17\%$ at $110$~GBd (black circle (FFE), square (VNLE)) and $120$ GBd (black circle (FFE), square (VNLE)), where the NBRs are $\approx290$ Gb/s and $\approx309$ Gb/s, respectively.

Considering equalization with linear FFE only, $110$~GBd \mbox{PS-PAM-8} performs~$>5$~times better than $100$~GBd uniform \mbox{PAM-8}.
However, $120$~GBd PS-PAM-8 performs only marginally better than $110$ GBd uniform PAM-8. The performance of PS-PAM is
mainly impaired by the BW limitation of the OE components. VNLE brings gain and, as discussed before, uniform PAM-8
benefits most (more than $7$ times) from nonlinear equalization due to its high susceptibility to nonlinearity. 

\subsubsection{AIR vs Symbol Rate} \label{ssubsec_AIR_vs_SymbolRate}
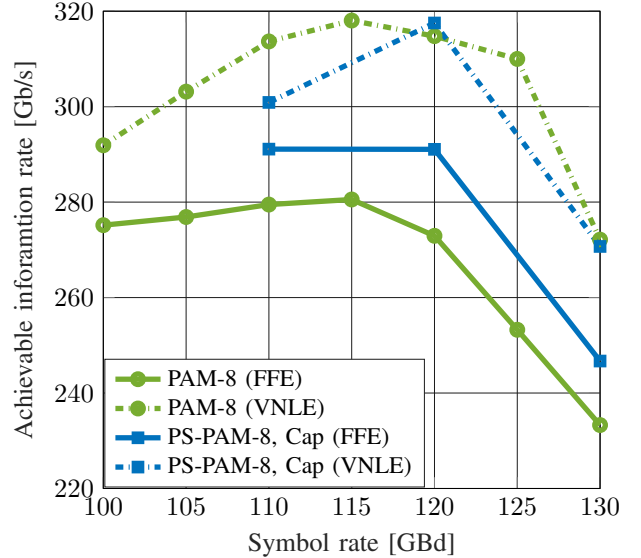
\begin{figure}[t]  
	\centering
%
%
\definecolor{mycolor1}{rgb}{0.46600,0.67400,0.18800}%
\definecolor{mycolor2}{rgb}{0.00000,0.44700,0.74100}%
\begin{tikzpicture}

\begin{axis}[%
width=2.6in,
height=2.5in,
at={(0.758in,0.562in)},
scale only axis,
xmin=100,
xmax=130,
xtick={100,105,110,115,120,125,130},
xlabel style={font=\color{white!15!black}},
xlabel={Symbol rate [GBd]},
ymin=220,
ymax=320,
ytick={200, 220, 240, 260, 280, 300, 320, 340, 360, 380},
ylabel style={font=\color{white!15!black}},
ylabel={Achievable inforamtion rate [Gb/s]},
axis background/.style={fill=white},
xmajorgrids,
ymajorgrids,
grid style={line width=.1pt, draw=black!90},
legend style={at={(0.002,0.002)}, {nodes={scale=0.9, transform shape}},anchor=south west, legend cell align=left, align=left, draw=white!15!black}
]
\addplot [color=mycolor1, line width=2.0pt, mark size=2.0pt, mark=o, mark options={solid, mycolor1}]
  table[row sep=crcr]{%
100	275.1674\\
105	276.8803\\
110	279.4854\\
115	280.5496\\
120	272.9399\\
125	253.2525\\
130	233.2876\\
};
\addlegendentry{PAM-8 (FFE)}

\addplot [color=mycolor1, dashdotted, line width=2.0pt, mark size=2.0pt, mark=o, mark options={solid, mycolor1}]
table[row sep=crcr]{%
100	291.9069796\\
105	303.1490\\
110	313.665554\\
115	318.0453\\
120	314.7758127\\
125	310.0013903\\
130	272.1384024\\
};
\addlegendentry{PAM-8 (VNLE)}

\addplot [color=mycolor2, line width=2.0pt, mark size=1.4pt, mark=square, mark options={solid, mycolor2}]
  table[row sep=crcr]{%
110	291.1031\\
120	291.0604\\
130	246.7034\\
};
\addlegendentry{PS-PAM-8, Cap (FFE)}

\addplot [color=mycolor2, dashdotted, line width=2.0pt, mark size=1.4pt, mark=square, mark options={solid, mycolor2}]
  table[row sep=crcr]{%
110	300.8864732\\
120	317.5309945\\
130	270.7095921\\
};
\addlegendentry{PS-PAM-8, Cap (VNLE)}

\end{axis}
\end{tikzpicture}%
	\caption{AIR of uniform and cap-shaped PS PAM-8.}
	\label{fig_AIRvsSymbolRate}
\end{figure}

In Fig. \ref{fig_AIRvsSymbolRate} we present the additional performance metric AIR, which is computed from the BER
results (Fig. \ref{fig_BERvsSymbolRate}) under the assumption of a binary symmetric channel (BSC). AIR indicates the possible rate of error-free data transmission with
capacity-achieving FEC. The AIR for HD is calculated in three steps, following \cite[Sec.~8.2.4]{bocherer2018principles}.
\begin{itemize} 
	\item Step 1: Compute the equivocation under bitwise HD by 	\begin{align}
		\textnormal{H}_2(p) = -p \cdot \log_2{(p)} - (1-p) \cdot \log_2{(1-p)}, \label{eq_entropy_BER}
	\end{align}
	where $\textnormal{H}_2$ is the binary entropy function and $p=$ BER.
	\item Step 2: Compute the achievable SE
	\begin{align}
		\textnormal{SE}_{\textnormal{HD}} = \left[\textnormal{H}(X) - m \cdot \textnormal{H}_2(\textnormal{BER})\right]^+, \label{eq_SE}
	\end{align}
	where $m$ is the number of bits per symbol, e.g, $m$ = $3$ for PAM-8 and $\textnormal{H}(X)$ is the entropy of the input ($3$ and $\approx2.77$ bits per symbol for uniform and PS PAM-8, respectively).
	\item Step 3: Compute AIR:
	\begin{align}
		\textnormal{AIR} = f_s \cdot \textnormal{SE}_{\textnormal{HD}},	\label{eq_AIR}
	\end{align}
	where $f_s$ is the symbol rate.
\end{itemize}

As shown in Fig. \ref{fig_AIRvsSymbolRate}, with linear FFE, the highest AIR, which is $\approx291$ Gb/s, is achieved
with $110$~GBd cap-shaped PS-PAM-8, whereas uniform PAM-8 achieves its peak AIR of $\approx280$~Gb/s at $115$~GBd. In the
presence of VNLE, uniform PAM-8 and PS PAM-8 achieve the same peak AIR of $\approx$~318~Gb/s at $115$~GBd and $120$~GBd,
respectively. We emphasize that PS-PAM-8 at $110$ GBd with FFE achieves the same performance ($\approx291$~Gb/s) of
$100$~GBd uniform PAM-8 with VNLE in terms of AIR.

\subsubsection{ONBR vs Symbol Rate} \label{ssubsec_ONBR_vs_SymbolRate}

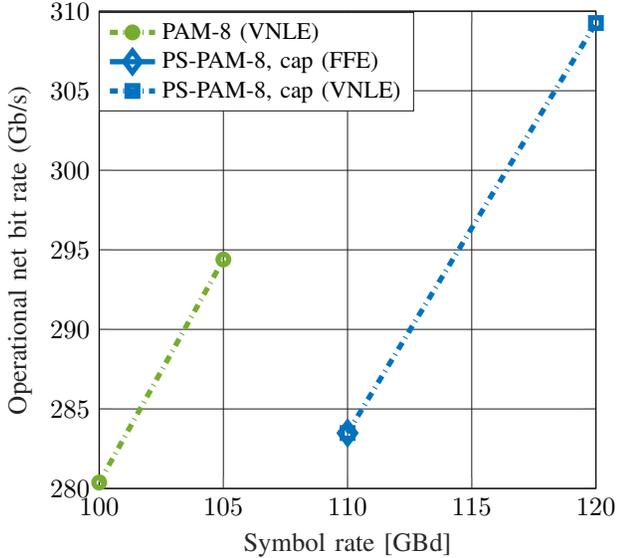
\begin{figure}[t] 
	\centering
%
%
\definecolor{mycolor1}{rgb}{0.46600,0.67400,0.18800}%
\definecolor{mycolor2}{rgb}{0.00000,0.44700,0.74100}%
\begin{tikzpicture}

\begin{axis}[%
width=2.6in,
height=2.5in,
at={(0.758in,0.562in)},
scale only axis,
unbounded coords=jump,
xmin=100,
xmax=120,
xlabel style={font=\color{white!15!black}},
xlabel={Symbol rate [GBd]},
ymin=280,
ymax=310,
ylabel style={font=\color{white!15!black}},
ylabel={Operational net bit rate (Gb/s)},
axis background/.style={fill=white},
xmajorgrids,
ymajorgrids,
grid style={line width=.1pt, draw=black!90},
legend style={at={(0.634,0.79)},{nodes={scale=0.9, transform shape}}, anchor=south east, legend cell align=left, align=left, draw=white!15!black}
]
\addplot [color=mycolor1, dashdotted, line width=2.0pt, mark size=2.0pt, mark=o, mark options={solid, mycolor1}]
  table[row sep=crcr]{%
100	280.3738\\
105	294.3925\\
110	nan\\
115	nan\\
120	nan\\
125	nan\\
130	nan\\
};
\addlegendentry{PAM-8 (VNLE)}

\addplot [color=mycolor2, line width=2.0pt, mark size=4pt, mark=diamond, mark options={solid, mycolor2}]
  table[row sep=crcr]{%
110	283.4852\\
120	nan\\
130	nan\\
};
\addlegendentry{PS-PAM-8, cap (FFE)}

\addplot [color=mycolor2, dashdotted, line width=2.0pt, mark size=2pt, mark=square, mark options={solid, mycolor2}]
  table[row sep=crcr]{%
110	283.4852\\
120	309.2566\\
130	nan\\
};
\addlegendentry{PS-PAM-8, cap (VNLE)}

\end{axis}

\end{tikzpicture}%
	\caption{ONBR of uniform and cap-shaped PS PAM-8.}
	\label{fig_ONBRvsSymbolrate}
\end{figure}

As further interpretation of the BER results (Fig. \ref{fig_BERvsSymbolRate}), we present in
Fig.~\ref{fig_ONBRvsSymbolrate} the net bit rate achieved with a practical FEC code, as a complement to
Fig.~\ref{fig_AIRvsSymbolRate}, which assumes capacity-achieving FEC. The ONBR is computed in two steps:
\begin{itemize} 
	\item Step 1: Compute the SE for the modulation formats whose BER lies below the considered threshold \hbox{($4.6\times10^{-3}$)}
	\begin{align}
		\textnormal{SE} = \left[\entop(X)-m(1-\rfec)\right]^+,
	\end{align}
	where $m$ is the number of bits per symbol, e.g., $m$ = $3$ for PAM-8, $\textnormal{H}(X)$ is the entropy of the input ($3$ and $\approx2.77$ bits per symbol for uniform and PS PAM-8, respectively), and $\rfec$ is the considered FEC OH, which is $7\%$.
	\item Step 2: Compute ONBR:
	\begin{align}
		\textnormal{ONBR} = f_s \cdot \textnormal{SE},	\label{eq_AIR}
	\end{align}
	where $f_s$ is the symbol rate.
\end{itemize}

With linear equalization only PS-PAM-8 (diamond marker) reaches the HD-FEC BER threshold, achieving an ONBR of
$\approx284$~Gb/s. On the other hand, with VNLE, both uniform (green dashed line with circle markers) and PS PAM-8 (blue
dashed line with square markers) reach the considered BER threshold with peak ONBRs of $\approx294$~Gb/s and
$\approx309$~Gb/s, respectively.

\section{Conclusion} \label{sec_conclusion}

In this paper, we have discussed \mbox{PS-PAM-8} for direct-detection systems and introduced a Gaussian order
(GO) parameter, which generalizes both the cap and cup-shaped MB distributions.

Cup-shaped distributions are expected to perform well in peak-power constrained (PPC) systems. However, since the
transfer function of a practical modulator, e.g. an EML, exhibits a gentle transition from the linear region towards
saturation, the considered system, differently from that considered in \cite{Thomas_PS_PAM_JLT21}, is not purely PPC. In
fact, we find that cap-shaped PS PAM-8 outperforms both uniform PAM-8 and cup-shaped PS PAM-8 by factors $>5$ and $>7$
in BER with linear equalization, respectively.

A theoretical analysis and a comparison of the signal histograms indicate that cap-shaped probabilistic shaping achieves
nonlinear gain at the cost of an increased symbol rate. In detail, a cap-shaped distribution lowers the occurrence
probability of the outer signal levels, which are less affected by saturation of the EML transfer curve. This allows an
increase of the driving voltage and results in a larger separation between adjacent modulated levels. It is important
to observe that the optimization of the driving voltage is fundamental to reveal the nonlinear gain of cap-shaped PAM.

We also investigated modified MB distributions with GOs $3.5$ and $5.0$ (besides GO $2.0$), which
further reduce the occurrence of the outer symbols. Although, as expected, a larger GO accommodates a higher
driving voltage, the additional BER improvement is marginal due to the severe bandwidth limitation of our experimental
setup.

Further, we computed the achievable information rate (AIR) on the binary symmetric channel established between the bit
mapper at the transmitter and the hard demapper at the receiver. In the case of linear equalization, on the considered
setup, the peak AIR for uniform PAM-8 and PS PAM-8 is $\approx280$~Gb/s and $\approx291$~Gb/s, respectively. In the
presence of nonlinear Volterra equalization, the nonlinear PS gain with respect to uniform PAM-8 is reduced, and both
modulation formats achieve a similar peak AIR of $\approx318$~Gb/s. Therefore, we conclude that the use of PS is a
low-complexity alternative to power-hungry nonlinear equalization. For example, we observe that \hbox{$110$~GBd}
\mbox{PS-PAM-8} with linear equalization reaches the same AIR as \hbox{$100$~GBd} uniform PAM-8 with advanced VNLE at the
expense of a $10\%$ increase in the symbol rate. Additionally, with practical FEC and linear equalization only
cap-shaped PS-PAM-8 achieves an operating net bit rate of $\approx284$~ Gb/s.
\vspace{-3mm} 
\section{Appendix} \label{sec_Appendix}

\begin{figure*}[h]
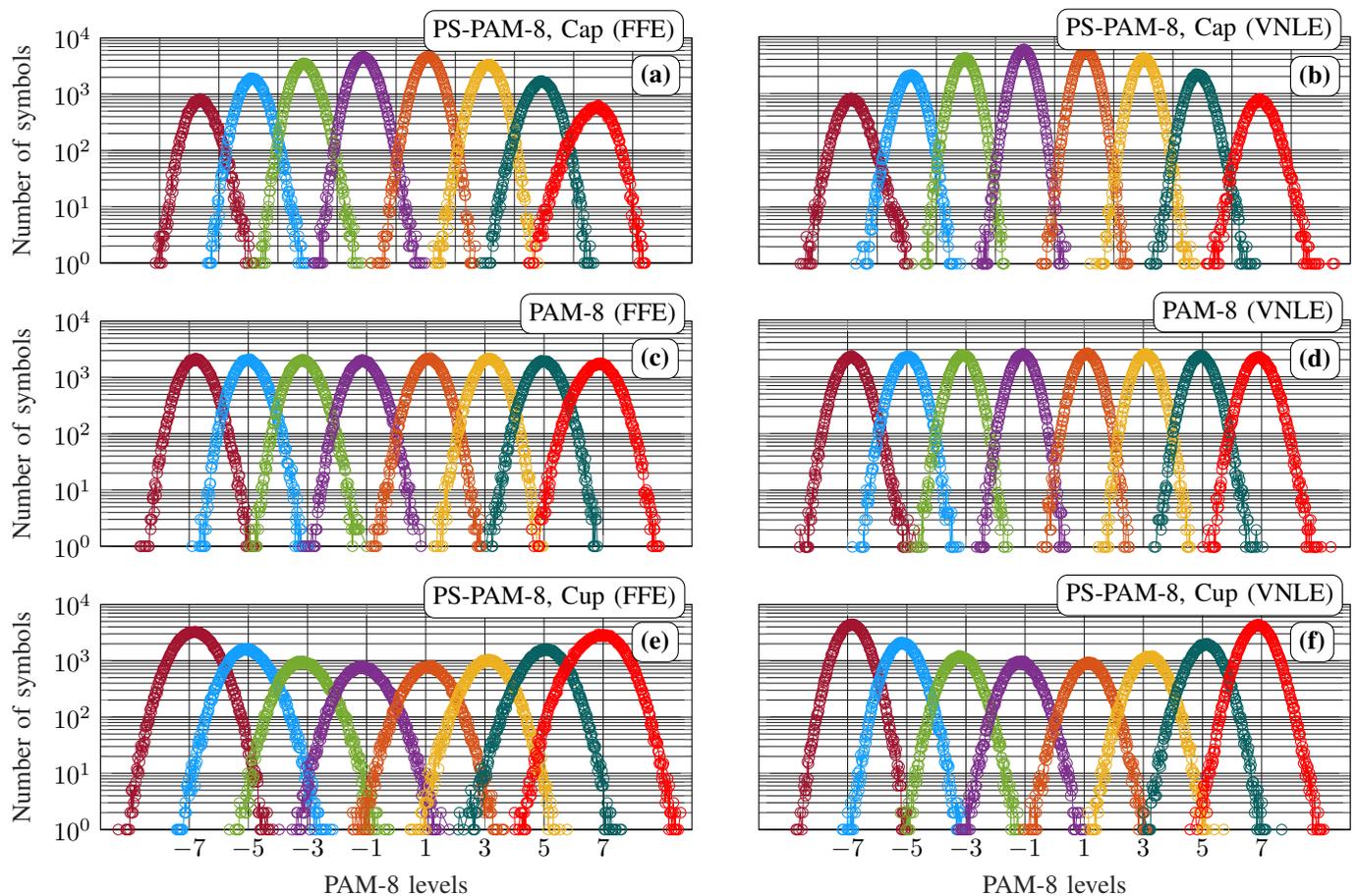

	\centering	
	\begin{subfigure}{0.45\textwidth}
		\centering		
		\input{figures/LevelDistributions_Tex/Cap_FFE.tex} 		
	\end{subfigure}\hfill
	\begin{subfigure}{0.45\textwidth}
		\centering		
		\input{figures/LevelDistributions_Tex/Cap_VNLE.tex} 		
	\end{subfigure}
	~
	\begin{subfigure}{0.45\textwidth}
		\centering
		\input{figures/LevelDistributions_Tex/PAM8_FFE.tex} 
	\end{subfigure}\hfill
	\begin{subfigure}{0.45\textwidth}
		\centering
		\input{figures/LevelDistributions_Tex/PAM8_VNLE.tex} 
	\end{subfigure}
	~
	\begin{subfigure}{0.45\textwidth}
		\centering
		\input{figures/LevelDistributions_Tex/Cup_FFE.tex} 
	\end{subfigure}\hfill
	\begin{subfigure}{0.45\textwidth}
		\centering
		\input{figures/LevelDistributions_Tex/Cup_VNLE.tex} 
	\end{subfigure}
	
	\caption{Level distributions after linear and nonlinear equalization for cap-shaped PS-PAM-8 are shown in (a) and
		(b), respectively. Similarly, the distributions for uniform PAM-8 and cup-shaped PS-PAM-8 are shown in (c), (d),
		(e) and (f), respectively.}
	\label{fig_LevelDistribustion_uniform_cap_cup}
	\vspace{-7mm}
\end{figure*}

An alternative visualization of Fig. \ref{fig_histogram_cap_or_cup} is presented in Fig.
\ref{fig_LevelDistribustion_uniform_cap_cup}, which shows the level distribution for each PAM-8 symbol after linear FFE
and nonlinear VNLE. The overlap among neighboring histograms contributes to the number of symbol errors. For cap-shaped
PS PAM-8 with FFE only (Fig. \ref{fig_LevelDistribustion_uniform_cap_cup} (a)) the overlap is minimized due to the high
driving voltage. Conversely, for cup-shaped \mbox{PS-PAM-8}, the limited driving voltage swing results in a massive overlap
among neighboring levels. An intermediate scenario is observed for uniform \mbox{PAM-8}. With VNLE the level distributions of
all formats are improved with uniform PAM-8 benefiting the most.

%
%
\bibliographystyle{IEEEtran}
\bibliography{Bibliography-sh,Bibliography-gb}
%
%
%
%
%

\end{document}